\documentstyle[12pt]{article}
\renewcommand{\^}{\hat}
\renewcommand{\dot}{\cdot}
\begin{document}
\begin{center}
{\bf Collinear N\'eel-type ordering in partially frustrated lattices}
\vskip 1cm
Uma Bhaumik and Indrani Bose 
\\Department of Physics \\Bose Institute
\\93/1, Acharya Prafulla Chandra Road
\\Calcutta-700 009, India.
\end{center}

\begin {abstract}
We consider two partially frustrated S = $\frac{1}{2}$ antiferromagnetic spin systems
on the triangular and pentagonal lattices.  In an elementary  plaquette of the 
two lattices, one bond has exchange interaction strength $\alpha$ ($\alpha\,\leq\,1$)
whereas all other bonds have exchange interaction strength unity. We show
 that for $\alpha$ less than a critical value $\alpha_{c}$, collinear
N\'eel-type ordering is possible in the ground state. The ground state energy
and the excitation spectrum have been determined using linear spin wave theory
based on the Holstein-Primakoff transformation.
\end{abstract}

P.A.C.S. Numbers: 75.10.Jm, 75.30.Ds

Frustrated spin sytems show a tendency to be magnetically disordered. 
Frustration may occur due to the presence of further neighbour interactions,
 besides the nearest-neighbour(NN) ones, as well as due to the topology of 
 the underlying lattice. A well-known example of the latter is the triangular lattice.
 The ground state of the Ising Antiferromagnet(AFM), defined on the triangular
 lattice, is highly degenerate with the entropy per site being a finite 
 quantity. The Ising AFM thus does not order at any temperature. In this 
 context, the question of interest is whether for a quantum spin Hamiltonian 
 long range magnetic order  can exist in the ground state. The quantum Heisenberg
 AFM has been widely studied on the triangular lattice \cite{Huse,Yoshika,
Deutscher, Wang,Jolicoeur} and there is now 
more or less a cosensus that AFM long range order(LRO) exists in the ground 
state. The ground state of the classical spin systems exhibits non-collinear 
$120^{0}$ ordering. The ordering is partially destroyed but still exists once 
the quantum nature of the spins is taken into account .

Frustrated lattices, in general, exhibit non-collinear N\'eel-type magnetic order 
whereas non-frustrated lattices like the square lattice exhibit collinear 
N\'eel-type order in the classical ground state. A partially frustrated lattice is 
obtained when not all the exchange interaction strengths along NN bonds have
equal values. In this case, the classical ground state, in a particular 
parameter regime, exhibits collinear N\'eel-type ordering similar to that in 
the case of a non-frustrated lattice. The ordering is not destroyed when quantum 
fluctuations are taken into account. Two examples of this will be given in this 
Brief Report, namely, those of the triangular and the pentagonal lattices.
The result for the classical ground state of the triangular lattice is already
known \cite{Zhitomirsky}. In the so called Row model, the strength of the unequel exchange 
interaction is $\alpha$  ($\alpha\leq1$) whereas all other exchange interaction
strengths have value unity (Fig.1). For $(\alpha\leq\alpha_{c})$, the ground state 
has collinear order whereas for $(\alpha>\alpha_{c})$, non-collinear spiral 
ordering occurs. This ordering becomes the $120^{o}$ ordering when $\alpha=1$.
For $\alpha\,\leq\,\alpha_{c}$, we determine the excitation spectrum as well as the quantal correction to the 
ground state energy using linear spin wave (LSW) theory  
based on the Holstein-Primakoff(HP) transformation. We next consider the case 
of the partially frustrated pentagonal lattice. The Ising AFM has been previously studied
on the pentagonal lattice \cite{Wolff,Waldor} and the ground state has been found to be disordered.
In this Brief Report, we study the spin $S=\frac{1}{2}$ HAFM on the pentagonal 
lattice for the first time. The elementary plaquette of a pentagonal lattice 
is a pentagon and hence the lattice is topologically frustrated. We study the 
model in a limited parameter regime, namely, the one in which collinear 
N\'eel-type order exists in the ground state. 

Consider a single triangular plaquette with three NN bonds. One of the bonds 
has exchange interaction strength $\alpha$, the other two have exchange 
interaction strengths unity. The three spins sit at the three vertices and
interact through NN interaction. If the spins are treated as classical 
vectors, it is easy to show that for $\alpha \leq \frac{1}{2}$, collinear 
N\'eel-type ordering is obtained in the ground state. Let 0, Q, $Q'$
be the respective orientations of the spins with respect to the z-axis. The 
first and third spins interact through exchange interaction of strength 
$\alpha$. In the collinear state, $Q=\pi$ and $Q'=0$. For $\alpha>0.5$,
$Q'=2Q$ in the ground state with $Q=cos^{-1}(-\frac{1}{2\alpha})$, i.e.,
a spiral ordering is obtained. We first consider the Row model (Fig.1) with 
$\alpha \leq \frac{1}{2}$. The sites belonging to alternate rows (shown 
by dashed lines) belong to one of the sublattices A or B. The interaction 
Hamiltonian is
\begin{equation}
H = \sum_{i\in A,B,\delta = \^{a_2}\, \^{a_3}}{\bf S_{i}\dot S_{i+\delta}}+ 
\alpha \sum_{i\in A,B,\delta=\^{a_{1}}}{\bf S_{i}\dot S_{i+ \delta}} 
\end{equation}
$\^a_{1}$, $\^a_{2}$, $\^a_{3}$ are the NN vectors in the horizontal and oblique 
directions respectively. The spins on the A(B) sublattice are pointing up(down).
This is the classical ground state. The foundation of spin wave theory is the 
assumption that AFM LRO exists in the ground state and the amplitude of 
zero-point motion produced by quantum fluctuation about the classical ordered
state is small. This assumption fails if quantal corrections diverge. The first 
step in the LSW theory is to transform the operators to the bosonic operators.
The HP transformations connecting the spin operators to the bosonic operators 
$a_j$,$b_j$'s are given in the lowest order by 
\begin{eqnarray}
S_{A_j}^{\dag}\,=\,\sqrt{2s}a_{j} \nonumber\\         S_{B_{j}}^{\dag}\,=\,\sqrt{2s}\,b_{j}^{\dag}\nonumber     \\
S_{A_{j}}^{z}\,=\,s-a_{j}^{\dag} a_{j}\nonumber\\      - S_{B_{j}}^{z}\,=\,s-b_{j}^{\dag} b_{j} 
\end{eqnarray}
where j denotes the lattice site and S is the magnitude of the spins. We will
ultimately consider the case of $S=\frac{1}{2}$. $S^{\dag}$, $S^{z}$ are the spin
raising operator and z component of the spin respectively.
One can similarly define the spin lowering operator $S_{j}^{-}$. We will not 
exhibit the different steps of LSW theory as these are standard \cite{Anderson,Kubo}.
The Hamiltonian(1) is expressed in terms of bosonic operators and then Fourier 
transformed. The Hamiltonian contains only quadratic operators and so can be 
diagonalized by the well-known Bogolyubov transformation. The diagonalized
Hamiltonian is given by 
\begin{eqnarray}
H = N\,S(S+1)\,(2\alpha-4) +S\sum_{k}\omega_{k} + S\sum_{k}\omega_{k}(c_{k}^{\dag} c_{k}
+ d_{k}^{\dag} d_{k})
\end{eqnarray}
where $c_{k}$,$d_{k}$ are the new transformed operators and N is the total 
number of sites. The excitation spectrum $\omega_{k}$ is given by
\begin{equation}
\omega_{k}=\sqrt{\left(1-\alpha \sin^{2}\frac{k\dot a_{1}}{2}\right)^{2}-\gamma_{k}^{2}}
\end{equation}
where
\begin{eqnarray}
\gamma_{k} = \frac{1}{z} \sum_{\delta=\^{a}_{2},\^{a}_{3}} e^{ik \dot \delta}
\end{eqnarray}
and z, the number of NNs along the oblique directions, is 4. Fig.2  shows the 
excitation  spectrum for $\alpha=0.3$ and S = $\frac{1}{2}$. The ground state 
energy $E_{g}$ (the sum of the first two terms in (3)) is given by $E_{g}$ = -0.6025.
One can verify that for $\alpha=0$, the excitation spectrum for the square lattice
is recovered. As long as $\alpha<\frac{1}{2}$, $\omega_{k}$ is positive for all 
momentum wave vectors. This shows that the choice of the starting 
ground state is correct. 

We next turn to the case of the pentagonal lattice. A cross-section of the 
lattice is shown in Fig 3. The lattice is a non-Bravais lattice and has 
two types of sites with coordination numbers 3 and 4 respectively.
The NN exchange interaction strengths along the solid lines are given by 
$\alpha$ ($\alpha \,\leq$ 1). Consider a single pentagonal plaquette .
The classical ground state for different values of $\alpha$ can be determined 
by minimizing the exchange interaction energy with respect to the spin orientation
angles. For $\alpha \, <0.4$, one finds that the collinear N\'eel state is the 
ground state. The question is whether this order survives once the quantal 
fluctuations are taken into account. Again, we perform LSW theory to get an 
answer. Fig.3 shows the six sublattices $(A_{i}\,B_{i}, \, i=1,2,3)$ corresponding 
to  the six inequivalent sites. The A(B) sublattices contain up(down)  spins.
The interaction Hamiltonian is given by
\begin{equation}
H=\sum_{<i \,j>} J_{i \, j}{\bf S_{i} \dot S_{j}}
\end{equation}
where <i \, j>  denotes NNs and $J_{i \, j}$=1 for NNs shown by solid lines
in  Fig.1. The NN vectors $\delta_{i}$'s are given by 
\begin{eqnarray}
{\bf \delta}_{1}=\^{x},\nonumber
\\{\bf \delta}_{2}=\^{y},\nonumber 
\\{\bf \delta}_{3}=\frac{1}{2}\^{x}+\frac{\sqrt{3}}{2}\^{y},\nonumber
\\{\bf \delta}_{4}=-\frac{1}{2} \^{x}+\frac{\sqrt{3}}{2}\^{y} 
\end {eqnarray}
and similarly for the vectors in the reverse directions. The HP transformations
connecting the spin operators  to bosonic operators $a_{i}$ and $b_{i}$ are given by (2). Each 
of the six  kinds of HP bosons forms a rectangular sublattice with dimension 
$(1,4+\sqrt{3})$. After Fourier transformation, the Hamiltonian(6) becomes
\begin{eqnarray}
H=H_{0}+H_{1}
\nonumber
\end{eqnarray}
where
\begin{eqnarray}
\\ H_{0}=NS(S+1)(-8+2\alpha)\nonumber\\
H_{1} =S\sum _{k}\psi_{k}^{\dag} \, M_{k} \psi
\\\psi_{k}^{\dag}=(a_{1k}^{\dag}\, a_{2k}^{\dag}\,a_{3k}^{\dag}\, b_{1k}\, b_{2k}\, b_{3k})
\end{eqnarray}
\begin{eqnarray}
M_k  = \left[\begin{array}{cc}
       Z_{1k} & Z_{2k}\\
       Z_{2k}^{\dag} & Z_{1k}
       \end{array}\right]\nonumber\\\nonumber\\\nonumber\\
\vspace{1cm}
Z_{1k}=\left[\begin{array}{ccc}
	 2-4\alpha\,\sin^2\frac{k\cdot \delta_1}{2} &\quad 0&\quad 0\\
	 0&\quad 3 & \quad 0\\
	 0&\quad  0 & \quad 3\end{array}\right]\nonumber\\\nonumber\\\nonumber\\
\vspace{1cm}
Z_{2k}=\left[\begin{array}{ccc}
	0&\quad x_1^{\star}&\quad x_2\\
       x_1&\quad 0&\quad x^{\star}\\
       x_2^{\star}&\quad x&\quad 0\end{array}\right]\nonumber\\\nonumber\\\nonumber\\
\vspace{1cm}
x_1=e^{ik\cdot\delta_1},\,\,\,
x_2=e^{ik\cdot\delta_2},\,\,\,
x=e^{ik\cdot\delta_3}+e^{ik\cdot\delta_4}
\end{eqnarray}
N is the total number of sites. Following  Jolicoeur and Le Guillou ,
we consider the generalised Bogolyubov transformation matrix T as follows
\begin{eqnarray}
\left[\begin{array}{c}
	\alpha(k)\\\beta(k)\end{array}\right]
= T \left[\begin{array}{c} a(k)\\b^{\dag}(k)\end{array}\right]
\end{eqnarray}
where $\alpha(k)$ is the column vector with three components $\alpha_{n}$(k)
(n=1,2,3) and similarly for $\beta^{ \dag}$(k). In order to satisfy the bosonic
commutation relations, T has to satisfy
\begin{eqnarray}
T^{-1}\,=\,\eta\,T\,\eta
\end{eqnarray}
where
\begin{eqnarray}
\eta=\left[\begin{array}{cc} I&\quad 0\\0 &\quad -I\end{array}\right]
\end{eqnarray}
 and I is the $3\times3$  identity matrix. Finally we have
 \begin{eqnarray}
H_1=S\sum_k[\alpha^{\dag}(k)\quad \beta(k)] \eta\,T[\eta\,M_k]\,T^{-1}
      \left[\begin{array}{c}\alpha(k)\\\beta^{\dag}(k)\end{array}\right]
 \end{eqnarray}
To diagonalize $H_{1}$ a suitable choice for T is
\begin{equation}
T^{-1}=(v^{1}\, v^{2}\, v^{3}\, w^{1}\, w^{2}\, w^{3})
\end{equation}
where $v^{n}$ and $w^{n}$ are the eigenvectors of $\eta M_{k}$, with corresponding 
eigenvalues ($\omega_{n}\,\,,-\omega_{n}$). The eigenvalues of $\eta M_{k}$  
occur in pairs with $\omega_{n}$>0. The matrix $\eta M_{k}$ can be diagonalized
numerically. We obtain  three excitation spectra each of which is doubly degenerate
corresponding to the two sublattices $A_{i},B_{i}$. The diagonalized Hamiltonian 
has the form
\begin{equation}
H=H_{0}+S\sum_{n=1}^{3} \omega_{n}(\alpha_{n}^{\dag}\alpha_{n}+\beta_{n}^{\dag}\beta_{n})
\end{equation}
\begin{equation}
H_{0}=NS(S+1)(-8+2\alpha) + S\sum_{n=1}^{3}\omega_{n}
\end{equation}
Fig.4 shows the three  excitation spectra for $\alpha=$0.27 and $S=\frac{1}{2}$.
The positivity of the spectra shows that the collinear N\'eel state with quantum
corrections is still the ground state. For $\alpha$ beyond the critical value
$\alpha_{c}$ $\approx $0.32 this is no longer so. The ground state energy $E_{g}$
is given by Eq.(18) and has the value $E_{g}$ =-0.4835 for 
$\alpha$=0.27.

For both the triangular and pentagonal lattices, there is a critical value 
$\alpha_{c}$ of $\alpha$ below which the classical ground state has collinear 
N\'eel-type order. As shown by LSW theory, the order is maintained, albeit with 
quantum corrections, when the quantum nature of the spins is taken into account.
Let us now consider the case $\alpha>\alpha_{c}$. For the triangular lattice, 
as mentioned before,  the ground state shows spiral ordering. For the pentagonal 
lattice, however, the problem is more difficult. Consider the case $\alpha$=1.
Calculations for an elementary plaquette show that in the classical ground state 
the difference in the successive spin orientation angles is $144^{0}$. Unlike
the case of the triangular lattice, the spin arrangement of a single pentagon 
cannot be repeated for the whole lattice. The determination of the classical ground
state structure thus becomes computationally more difficult. The parameter region
$\alpha>\alpha_{c}$ for the pentagonal lattice has not been studied as yet.
The value of $\alpha_{c}$ can be taken as a measure of frustration in a spin system.
The values for the square, triangular and pentagonal lattices are 
$\alpha_{c}$ =1, 0.5 and $\sim$0.32 respectively. Thus the pentagonal lattice 
appears to be more frustrated than the triangular lattice. Several studies have 
been undertaken in the recent past to understand the effect of topological frustration
on ground state properties. In two dimensions, the most studied cases are 
those of the triangular and Kagom\'e lattices.
Frustration, in general, leads to more disorder in the ground state. The interest
in spin-disordered states has arisen in connection with high temperature superconductivity.
The pentagonal lattice is another topologically frustrated lattice which has not 
been studied so far in the context of quantum spin systems. The present study 
constitutes a small begining  which  will hopefully lead to more exhaustive studies in future.

\section*{Acknowledgement}
The Authors thank Sitabhra Sinha and Asimkumar Ghosh for computational help.

\newpage
\section*{Figure Captions}
\begin{description}
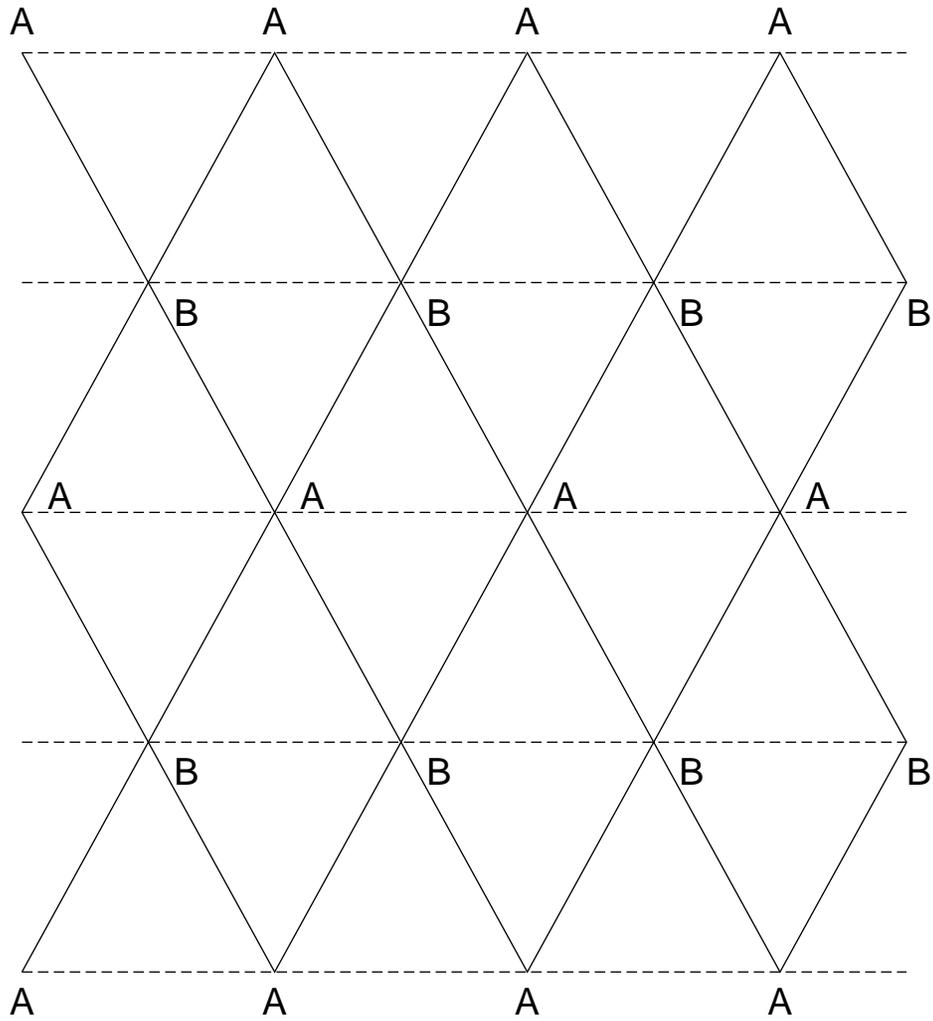
\item[Fig.1] The partially frustrated Row model defined on the tringular lattice.
 A and B denote the two sublattices. The dashed bonds have exchange interaction
strength $\alpha(\alpha\leq1)$, all other bonds (solid lines) have exchange 
interaction strength unity.

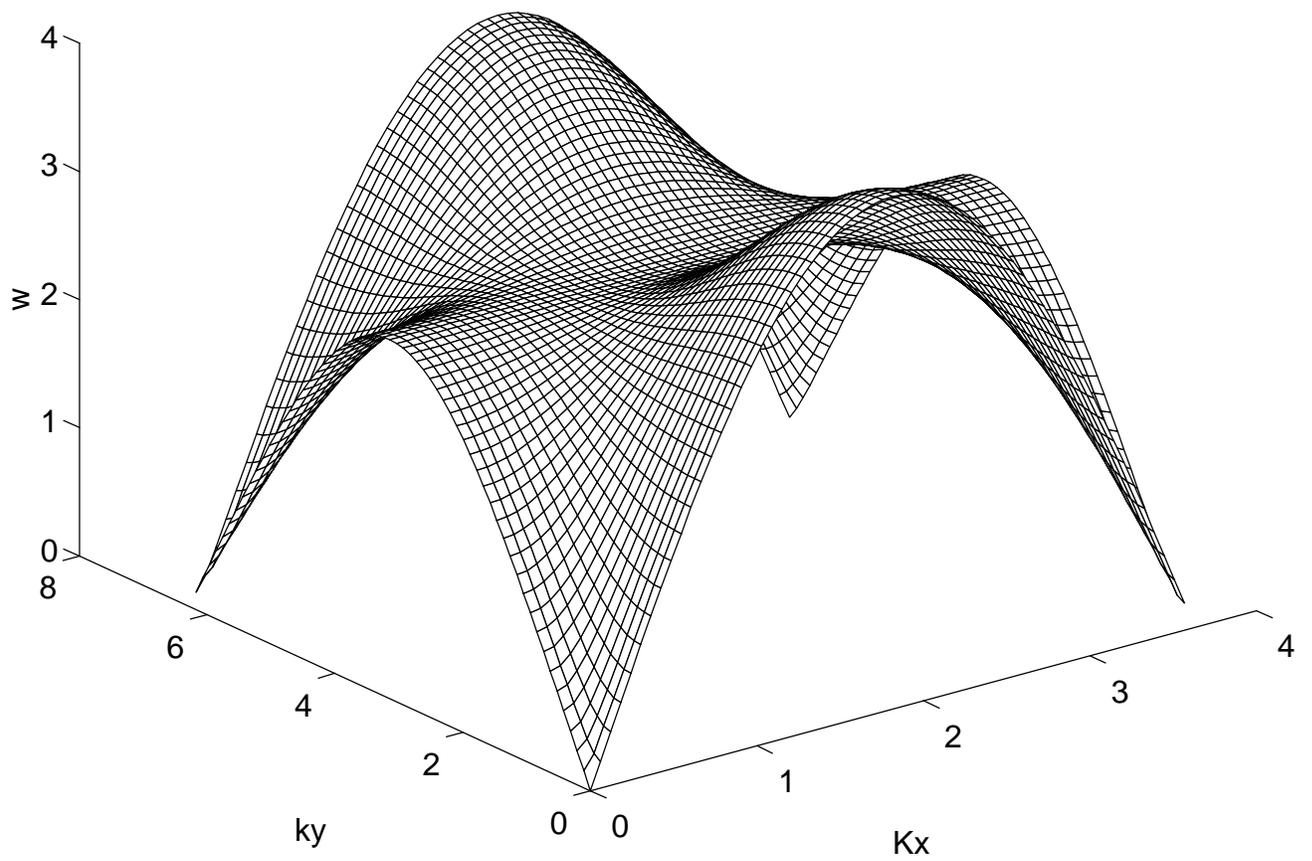
\item[Fig.2] Excitation spectrum of the Row Model for $\alpha$=0.3.

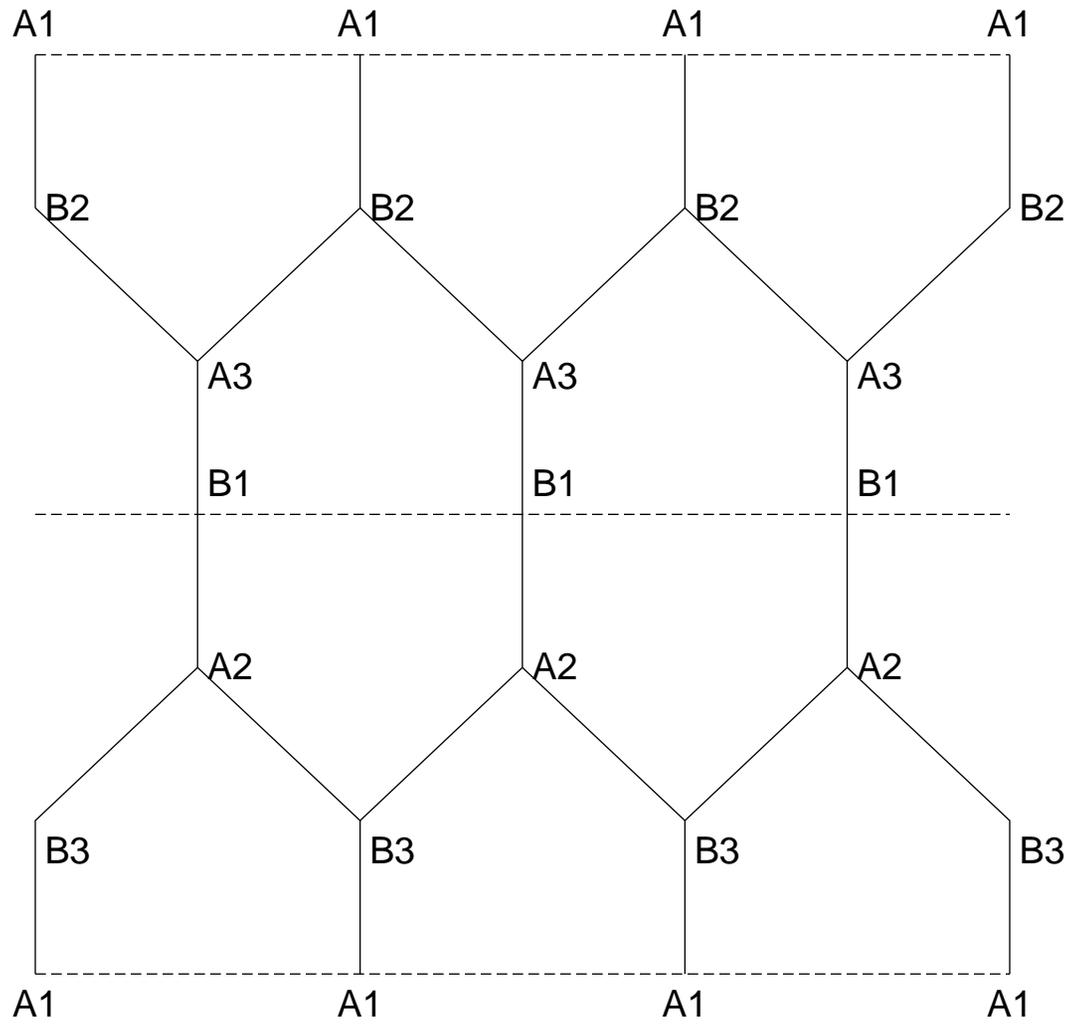
\item[Fig.3] The partially frustrated pentagonal lattice. The six sublattices
are $A_{i}$ and $B_{i}$ (i=1,2,3). The dashed bonds have exchange interaction 
strength $\alpha(\alpha<1)$, all other bonds (solid lines) have exchange 
interaction strength unity.

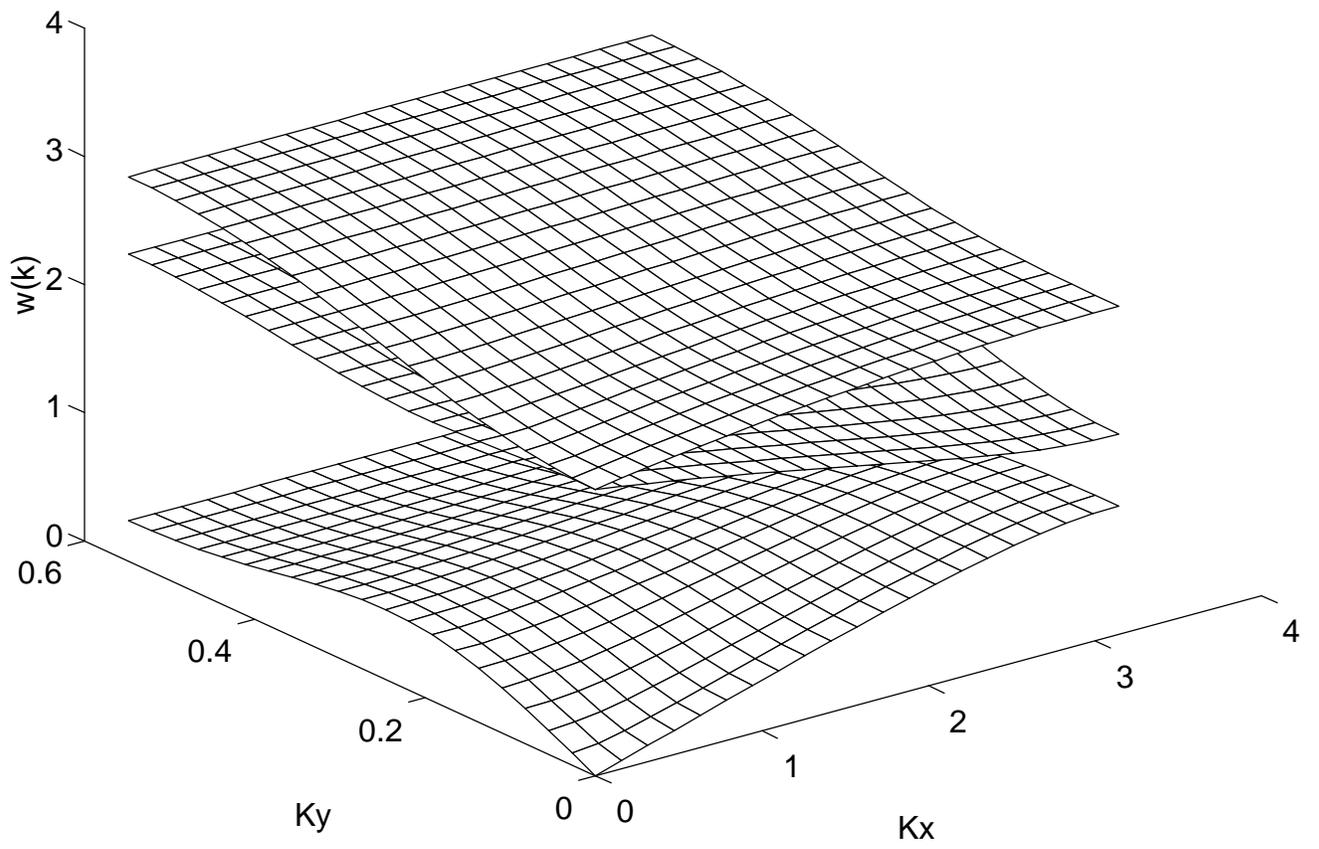
\item[Fig.4] Excitation spectra for the pentagonal lattice for $\alpha$=0.27.
\end{description}

\end{document}